\documentstyle[10pt,aasnewpp,psfig]{article}

\lefthead{Laine et al.}
\righthead{CO Distribution and Kinematics in NGC~7479}

\begin{document}

\twocolumn

\title{CO Distribution and Kinematics Along the Bar in the Strongly Barred 
Spiral NGC~7479}

\author{S. Laine}
\affil{Department of Physical Sciences,
University of Hertfordshire, College Lane, Hatfield, AL10~9AB,
United Kingdom, seppo@star.herts.ac.uk}

\author{J. D. P. Kenney}
\affil{Astronomy Department, Yale University, P.O. Box 208101,
        New Haven, CT 06520-8101, kenney@astro.yale.edu}
        
\author{M. S. Yun}
\affil{National Radio Astronomy Observatory, 1003 Lopezville Road,
        Socorro, NM 87801-0387, myun@nrao.edu}
        
\and

\author{S. T. Gottesman}
\affil{Department of Astronomy, P.O. Box 112055, University of 
        Florida, Gainesville, FL~32611-2055, gott@astro.ufl.edu}

\begin{abstract}
We report on the 2\farcs 5 (400 pc) resolution CO (J = 1 $\rightarrow$ 0) 
observations covering the whole length of the bar in the 
strongly barred late-type spiral galaxy NGC~7479. CO emission is detected
only along a dust lane that traverses the whole length of the bar,
including the nucleus. The emission is strongest in the nucleus. The 
distribution of emission is clumpy along the bar outside the nucleus,
and consists of gas complexes that are unlikely to be gravitationally
bound.
The CO kinematics within the bar consist of two separate components.
A kinematically distinct circumnuclear disk, $<500$~pc in
diameter, is undergoing predominantly circular motion with a maximum
rotational velocity of 245~km~s$^{-1}$ at a radius of 1\arcsec~(160~pc).
The CO-emitting gas in the bar outside the circumnuclear disk has
substantial noncircular motions which are consistent with a large radial
velocity component, directed inwards.
The CO emission has a large velocity gradient
across the bar dust lane, ranging from 
0.5 to 1.9~km~s$^{-1}$~pc$^{-1}$ after correcting for inclination and the 
projected velocity change across the dust lane is as high as 200 km~s$^{-1}$. 
This sharp velocity gradient is consistent with a shock
front at the location of the bar dust lane. 
A comparison of H$\alpha$ and CO kinematics across the dust
lane shows that although the H$\alpha$
emission is often observed both upstream and downstream from the dust lane, the
CO emission is observed only where the velocity gradient is large. 
We also compare the observations with hydrodynamic models and discuss star 
formation along the bar.
\end{abstract}


\section{Introduction}

To investigate the various factors that determine the molecular gas morphology
and kinematics in the bars of spiral galaxies we need to obtain observations 
of galaxies
which have molecular gas in both the nucleus and the bar. NGC~7479 is one of
the few galaxies which have abundant molecular gas along most of the bar
and in the nucleus (\cite{qui95}). It is an isolated, large, strongly barred 
SBc galaxy at a 
moderate distance (32 Mpc, assuming $H$$_{0}$~=~75~km~s$^{-1}$, a 
heliocentric velocity of $V$~=~2371 km~s$^{-1}$ and no correction
for the local group velocities; 1\arcsec~= 160~pc). 
It has both abundant atomic gas (\cite{lai98a}) 
and molecular gas (\cite{you91a}), and it has been classified as a starburst
galaxy based on its large and centrally concentrated 10 $\mu$m flux
(\cite{dev89}).

We have obtained new, high resolution (synthesized beam = 
2\farcs 7$\times$2\farcs 1, position angle~= $-85\arcdeg$) 2.6 mm CO
(J = 1 $\rightarrow$ 0) observations
of NGC~7479, and investigate the gas kinematics and distribution in the
unusually gas-rich bar. We have detected a strong central 
molecular gas component and weaker
emission along the bar, closely following the leading dust lane, as previously 
shown by the lower resolution (7\arcsec) CO maps (\cite{qui95}).
Our new data have greater sensitivity and spatial resolution. The emphasis 
in this paper is on the interesting CO
kinematics in both the nuclear area and in the offset gas/dust lane along the 
bar. 
Other papers in this series address the distribution and kinematics of the 
neutral, atomic
hydrogen gas in NGC~7479 (\cite{lai98a}), the stellar bar pattern 
speed of NGC~7479 (Laine, Shlosman, \& Heller 1998), and the minor merger model
for NGC~7479 (\cite{lai98c}).

\section{Observations}

The new CO observations were made at the Owens Valley Radio
Observatory (OVRO) Millimeter Array.\footnote{The Owens Valley Radio 
Observatory
Millimeter array is operated by the California Institute of Technology
with support from the National Science Foundation.} 
NGC~7479 was observed in three different configurations of the array.
The observations were made on
1994 October 7, November 23--24, and 1995 January 18-19. The total 
integration times at
each of the three observed positions and other observing parameters are
given in Table~\ref{tbl-1}. The total single sideband temperatures,
including the effects of the atmosphere, were typically 350--500 K. 
Variations in the sky opacity and receiver gain were 
corrected through measurements of an ambient temperature chopper wheel.
The pointing was checked at the beginning of each track, and the strong 
quasar 3C~454.3 was observed every 19 minutes to serve as a bandpass and 
gain calibrator, as well as a secondary flux calibrator.

\begin{table*}
 \begin{minipage}{175mm}
 \begin{center}
 \caption{Observing parameters at OVRO. \label{tbl-1}} 
 \begin{tabular}{@{}lc}
 \tableline
 \tableline
  Parameter & Value \\
 \tableline
 Pointing center 1 & R.A. 23$^{\rm h}$ 02$^{\rm m}$ 25\fs 99 \\
                   & Dec. 12\arcdeg~02\arcmin~24\farcs 2 \\
 Pointing center 2 & R.A. 23$^{\rm h}$ 02$^{\rm m}$ 26\fs 30 \\
                   & Dec. 12\arcdeg~03\arcmin~09\farcs 0 \\
 Pointing center 3 & R.A. 23$^{\rm h}$ 02$^{\rm m}$ 26\fs 61 \\
                   & Dec. 12\arcdeg~03\arcmin~53\farcs 7 \\                                    
 Central frequency & 114.37 GHz \\
 Central velocity (LSR) & 2340 km~s$^{-1}$ \\
 Configurations used & A, B, C \\
 Total number of antennas & A=5, B=6, C=6 \\
 Field of view & 63\arcsec \\
 Spatial scales sampled\tablenotemark{a} & 2\farcs 5 -- 50\arcsec \\
 Baseline lengths & 4.0 -- 92.7 k$\lambda$ \\
 Integration time (min), p.c. 1 & 350 \\
 p.c. 2 & 360 \\
 p.c. 3 & 360 \\
 Absolute flux calibrator & Neptune, $T_{B}$ = 114 K \\
 Secondary flux calibrator & 3C454.3 \\
 Velocity resolution & 10.4 km~s$^{-1}$ \\
 Total observed bandwidth & 464 MHz = 1206.4 km~s$^{-1}$ \\
 \tableline
 \end{tabular}
 \tablenotetext{a}{See text in \S~3.1~for more information.}
 \end{center}
 \end{minipage}
\end{table*}

\begin{figure*}
\psfig{figure=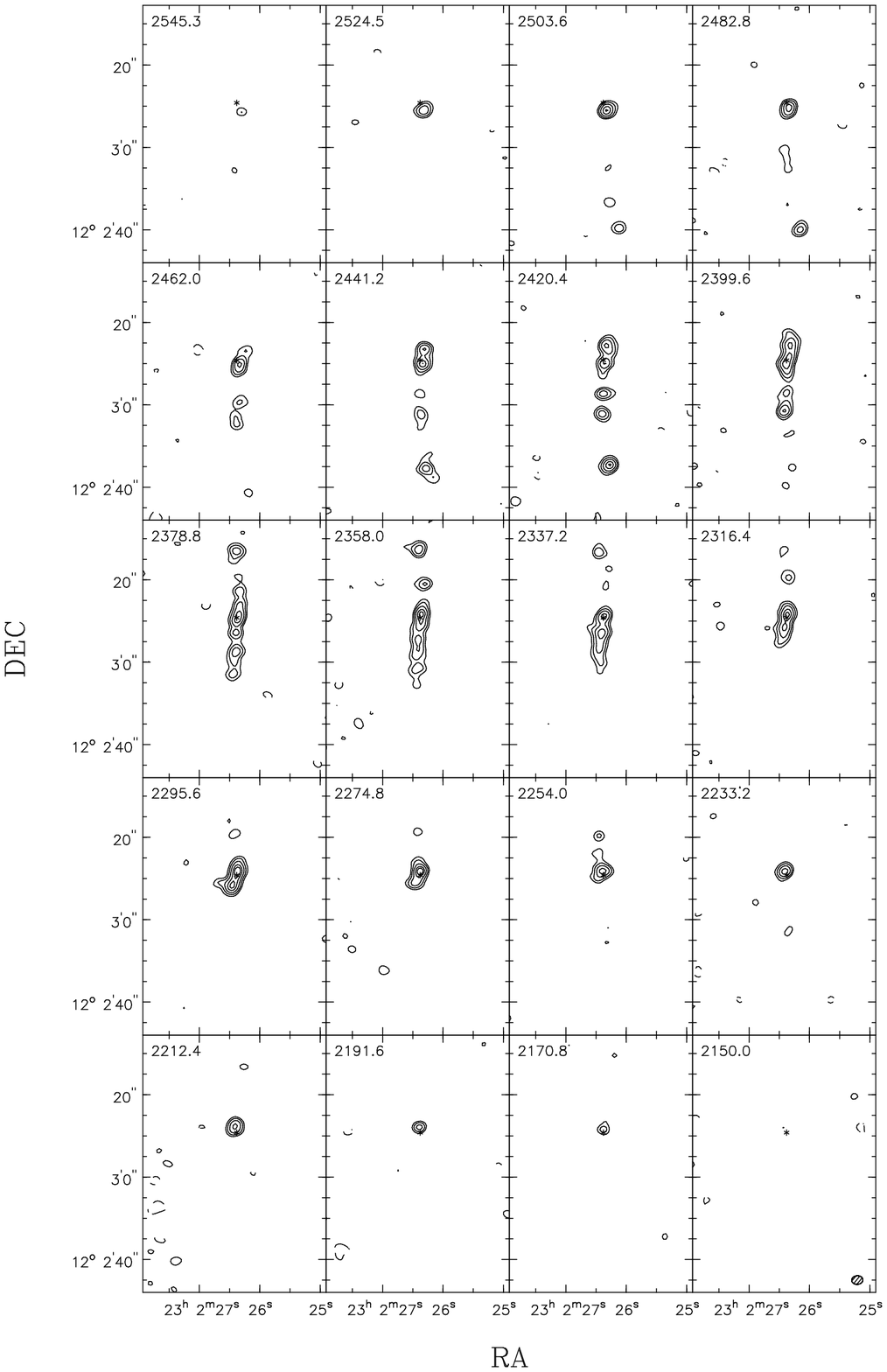,height=7.0in}
\caption{The 20~km~s$^{-1}$ channel maps of the naturally 
weighted CO cube. 
The contour levels are at ($-3$, 3, 5, 8, 12, 15) $\times$~0.021~Jy/beam. The star 
marks the $K$-band nucleus. The velocity at the center of each channel is given 
in the upper left of each frame and 
the beam (2\farcs 7 $\times$ 2\farcs 1) at the bottom right of the last 
frame. Note the strong emission near 
the nucleus in most of the channels,
due to the circularly rotating circumnuclear disk. \label{fig1}}
\end{figure*}

The calibration followed standard procedures. After correcting for the 
atmospheric and 
instrumental gain variations, we  
produced channel 
and integrated maps of the CO emission. For mapping the visibility data,
natural weighting was used to achieve the highest sensitivity.  We used 
15 mJy beam$^{-1}$ ($1.5\sigma$) as the
stopping criterion for the CLEAN iterations. The map parameters are 
given in Table~\ref{tbl-2}. We combined the 
three observed positions into a mosaic image, 
correcting for the attenuation of the primary beam.
The velocity-smoothed channel maps of the CO emission are shown in
Figure~\ref{fig1}. The measured line of sight CO velocities quoted in this
paper are with respect to the LSR unless otherwise stated.
The mapped regions have been drawn with great circles in
Figure~\ref{fig2}, which shows the integrated flux map. 

We also present data from H$\alpha$ Fabry--Perot
observations, kindly given to us by Drs. Stuart Vogel and Michael Regan. These
observations were made with the Maryland--Caltech Fabry--Perot
Spectrometer (see, e.g. Vogel et al. 1995) attached to the Cassegrain
focus of the 1.5 m telescope at Palomar Observatory. The data were
obtained
on 1994 September 29--30. 40 exposures were taken, each with a 500 second 
integration time and a pixel scale of 1\farcs 88~pixel$^{-1}$. To improve
the signal to noise ratio, these
data have been smoothed to 3\farcs 6 spatial resolution. The velocity planes 
are separated by 12.1 km~s$^{-1}$. The velocity uncertainty (rms) is estimated
as 1--2 km~s$^{-1}$. The data reduction procedure followed that
of Regan et al. (1996), who give a more detailed description.

\section{Results}

\subsection{CO Flux and Distribution}

\begin{figure}
\psfig{figure=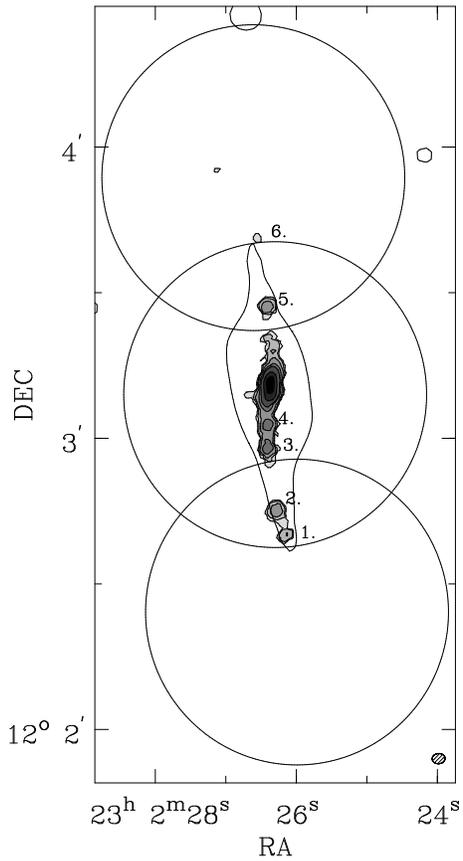,height=4.5in}
\caption{The naturally weighted integrated map of
$^{12}$CO(1--0) emission.
The contour and gray scale levels are (4, 8, 16,
32, 48, 96, 144) $\times$ 10$^{21}$ H$_{2}$
molecules cm$^{-2}$. The contour around the CO emission is the
17.1 mag arcsec$^{-2}$ level of a $K$-band image (Quillen et al. 1995), 
showing the location of the
stellar bar. The PA of the stellar bar and that of
the gaseous bar are different, since the gas is in a leading dust lane. The 
beam is shown at the bottom right corner of the map. The main CO clumps
have been numbered. The beams around the three observed positions have been 
drawn with large circles. \label{fig2}}
\end{figure}

\begin{figure}
\psfig{figure=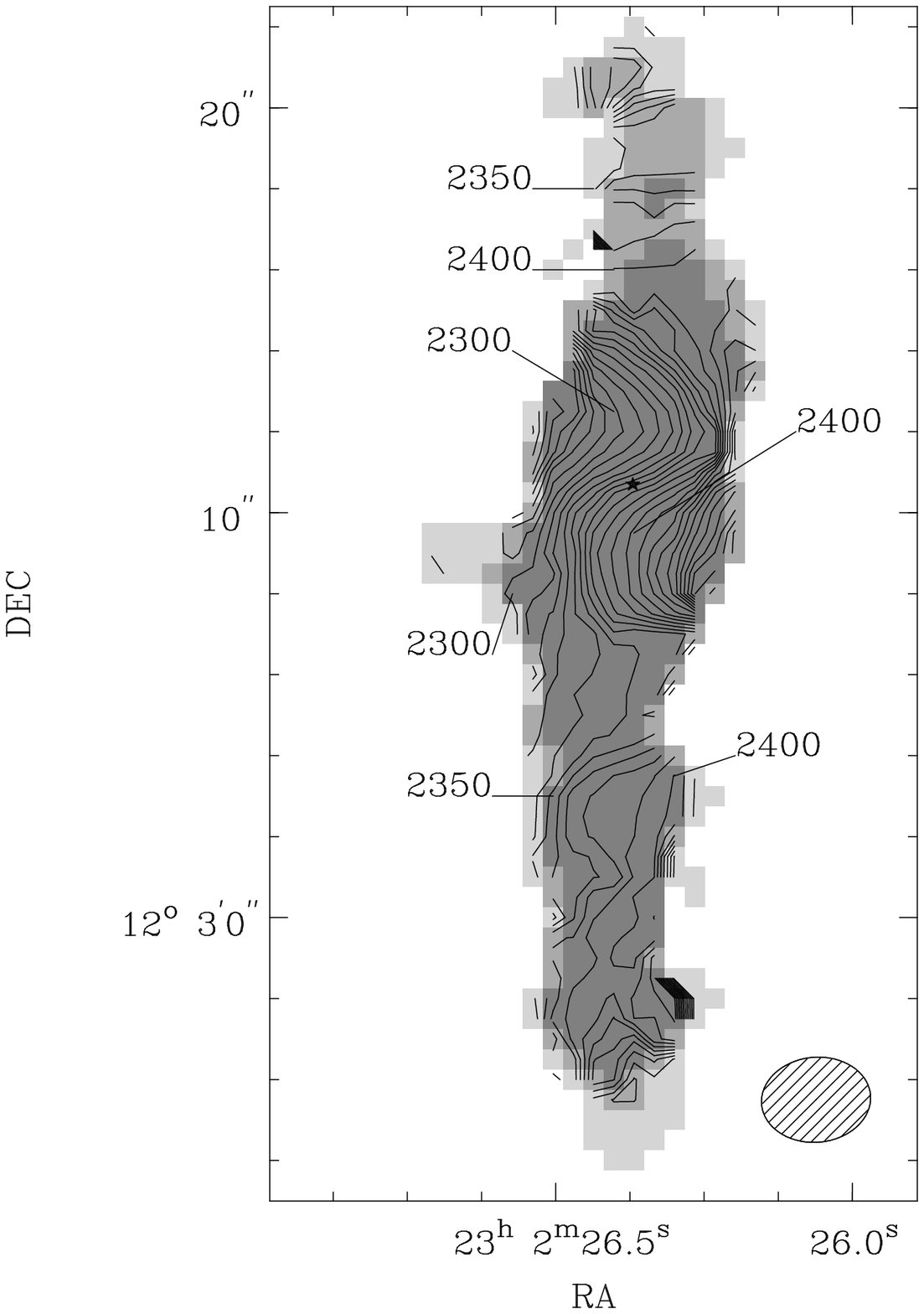,height=4.0in}
\caption{A magnified image of the central CO velocity
field superposed on the integrated CO emission map. The CO emission peak 
is marked with a star in the nucleus. The gray scale levels 
have been chosen to show only the brightest CO emission. The isovelocity 
contours in this map do not accurately reflect
the interesting gas kinematics in some regions.
The Z-shaped contours near the center occur where the inwardly flowing
bar gas meets the kinematically distinct circumnuclear disk.
These two distinct components are better displayed in Fig.~\ref{fig7}.
The large velocity gradient across the bar dust lane, which we 
have marginally resolved, is better displayed in Fig.~\ref{fig8}. \label{fig3}}
\end{figure}

The integrated $^{12}$CO (1--0) flux map of the whole mapped area and the 
mean velocity field of the central gas complex are 
displayed in Figures~\ref{fig2} and \ref{fig3}, respectively. The unit of the integrated
flux map has been converted to $N$(H$_{2}$) cm$^{-2}$ with the ``standard
conversion factor'' (\cite{blo86}; \cite{sco87}; \cite{ken89}), 

\begin{equation}
N({\rm H_2})~{\rm cm}^{-2}~=3.0 \times 10^{20}\int T_{\rm B}({\rm CO}){\rm
d}v~{\rm K}~{\rm
km}~{\rm s}^{-1}.
\end{equation}

\begin{figure}
\psfig{figure=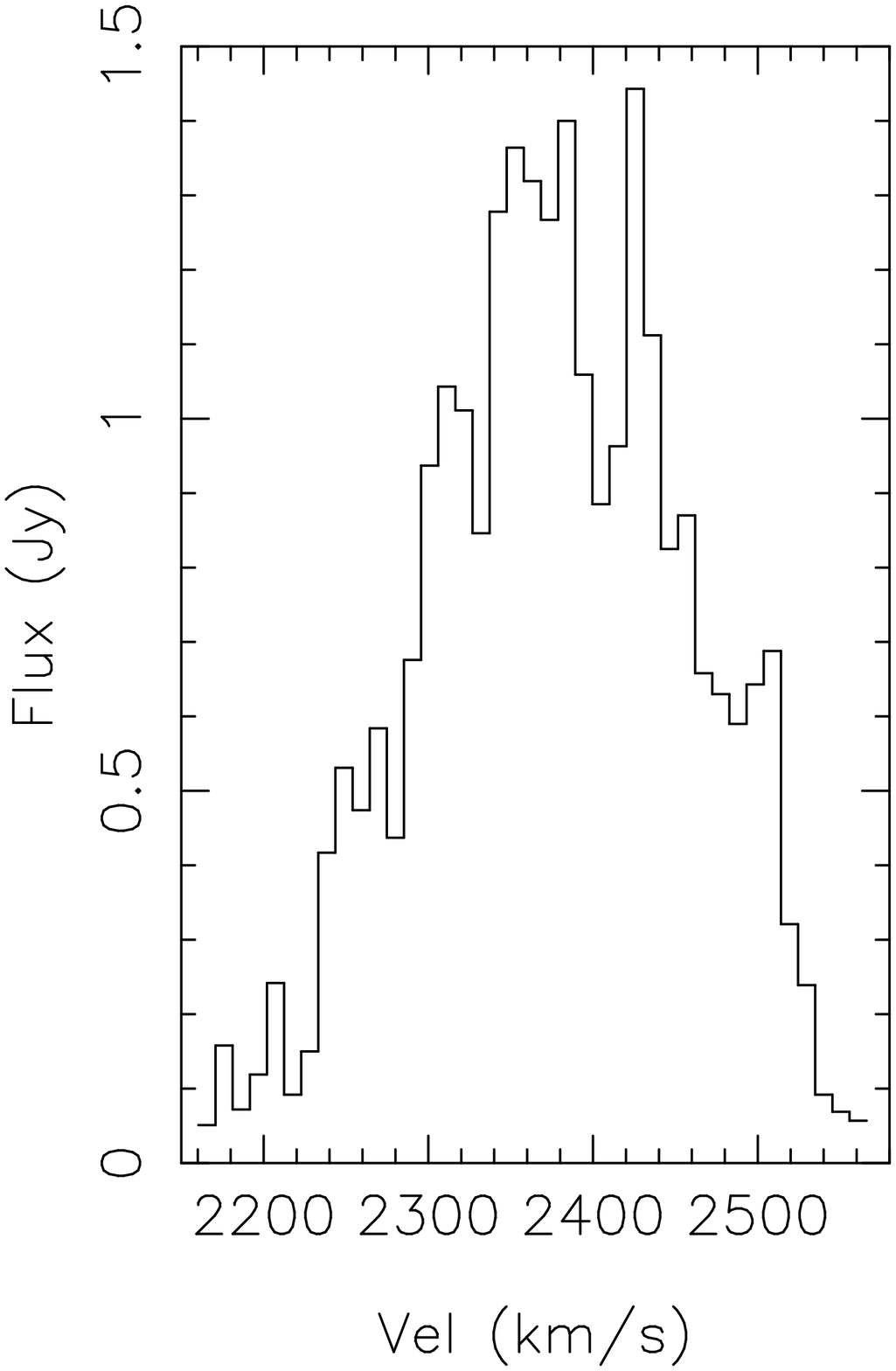,height=2.0in}
\caption{The spectrum of the CO emission detected with the 
OVRO millimeter interferometer. The channel fluxes have uncertainties of 
the order of 5\%. \label{fig4}}
\end{figure}

The CO flux spectrum is plotted 
in Figure~\ref{fig4}. The total integrated
flux is 
266$\pm$40 Jy km~s$^{-1}$,
where the stated uncertainty is the rms value in the measured fluxes. 
In addition, systematic uncertainties in the flux of the secondary calibrator 
($15\%$) can lead to a systematic
over- or underestimation of the fluxes. For comparison, the total, integrated 
single dish CO flux for the whole galaxy (4\arcmin$\times$3\arcmin) is 
1050$\pm$200~Jy~km~s$^{-1}$ (\cite{you95}), but for the central 
45\arcsec~only, we estimate a CO flux of 430$\pm$65~Jy km~s$^{-1}$
(\cite{ken96}).

At least $62\%\pm20\%$ of the flux in the central zone of the galaxy 
has been detected by our high resolution
OVRO interferometer observations. Our interferometer observations have
some sensitivity to structures up to 50$\arcsec$ in size, although the
sensitivity starts to decrease for structures with sizes larger than half
the field of view (structures with sizes $>$~30$\arcsec$).
The ``missing'' flux could be associated
with large scale structures ($>$~30\arcsec) for which our interferometer
measurements are not fully sensitive. However, it is unlikely that a
single velocity channel (width = 10~km~s$^{-1}$) would have continuous CO
emission structures this large. More likely, our brightness temperature 
sensitivity (rms = 0.2~K, corresponding to an H$_{2}$ column density of about
10$^{22}$ molecules~cm$^{-2}$ averaged over the beam, using the standard conversion
factor) is not good enough to detect the emission from regions with low
beam-averaged CO surface densities. We will keep these limitations in mind when
discussing the implications of our observations in \S~4.

\begin{table*}
 \begin{minipage}{175mm}
 \begin{center}
 \caption{Parameters of the $^{12}$CO channel maps. \label{tbl-2}}
 \begin{tabular}{@{}lc}
 \tableline
 \tableline
 Parameter & Value \\
 \tableline
 Size of synthesized beam & $2\farcs 7 \times 2\farcs 1$ \\
 Size of beam in parsecs\tablenotemark{a} & $262 \times 204$ \\
 PA of synthesized beam & -85\arcdeg \\
 Weighting used to make maps & Natural \\
 Rms noise in channel maps & 0.011 Jy beam$^{-1}$=0.19 K \\
 Theor. rms noise in channel maps & 0.0115 Jy beam$^{-1}$=0.20 K \\
 Peak brightness temperature & 3.83 K \\
 Radio continuum flux density\tablenotemark{b} & $<$ 4 mJy \\
 \tableline
 \end{tabular}
 \tablenotetext{a}{Assuming a distance of 32 Mpc.}
 \tablenotetext{b}{A 5$\sigma$ upper limit is given for a point source.}
 \end{center}
  \end{minipage}
\end{table*}

Fig.~\ref{fig2} reveals that all the detected CO
emission comes from a nearly linear ridge which is offset towards
the leading edges of the bar (assuming the spiral arms are
trailing). This ridge has a full-width-half-maximum (FWHM) value similar 
to the width
of the beam (2\farcs 7), indicating that the gas/dust lane is unresolved. 
The dust lane, as measured from 0\farcs 8 resolution
optical and near-infrared images, has a width of about 1\farcs 5. Therefore,
the high density molecular gas ridge does not have a width larger than that of the
dust lane. The 
CO emission intensity has a strong peak in the nucleus,
but overall the emission crosses the nucleus smoothly
along position angle (PA) $-10\arcdeg$, connecting the two linear, offset
parts of the bar gas/dust ridge.

Applying the standard CO--to--H$_{2}$
conversion factor and multiplying the resulting masses by 1.35 to account 
for helium, the molecular mass within the central 1\arcsec~(160 pc) radius is 
$3.3~\times$ 10$^{8}$ $M_{\sun}$ and the mean gas mass surface density
in this area is 4.1~$\times$~10$^{3}$~$M_{\sun}$~pc$^{-2}$. This should be 
compared
with the galactic center molecular gas mass of 3~$\times$~10$^{7}$~$M_{\sun}$
within the central 500 pc (Dahmen et al. 1998) that has an average
molecular gas surface
density of 150~$M_{\sun}$~pc$^{-2}$. The corresponding hydrogen column
density in the nucleus of NGC~7479 is about 
1.6~$\times$~10$^{23}$~cm$^{-2}$ which implies a beam-smoothed
optical extinction (A$_{\rm V}$) value of almost a hundred magnitudes. Extinction
will be further discussed
when addressing the star formation rates in \S~4.2. The molecular gas mass
in the central 7\arcsec~diameter region is about 2~$\times$10$^{9}$~$M_{\sun}$ 
and the total molecular mass derived from our observations is 
4~$\times$~10$^{9}$~$M_{\sun}$. Roughly half
(2~$\times$~10$^{9}$~$M_{\sun}$) of the detected flux (and
mass) lies along the bar outside the central 7\arcsec~diameter region. 

The detected CO emission along the bar is highly clumpy. 
In the integrated flux map (Fig.~\ref{fig2}) 
at least six separate clumps in addition to the nuclear gas complex can be
identified. 
The masses of the clumps along the bar range from
$9 \times 10^{6}$~$M_{\sun}$ to $2 \times$ 10$^{8}$~$M_{\sun}$. However,
the use of the standard CO-to-H$_{2}$ conversion factor may not be justified
in strongly shocked gas along the bar. The temperature and density are
likely to be higher in the shock, although these effects tend to offset
each other. If the clouds in the shock
are not gravitationally bound (see \S~4.1.2) their large line
widths cause the standard
CO--to--H$_{2}$ conversion to overestimate the H$_{2}$ mass. Shock
chemistry can also change the conversion factor. The lower limit for the
conversion factor is obtained for optically thin CO emission which has
a conversion factor 20 times smaller than the standard factor (cf.
Bryant \& Scoville 1996). Overall, it is
difficult to ascertain how the conversion factor would change in bar shocks.

The sizes of the clumps along the bar are too large to be
individual self-gravitating giant molecular clouds but their sizes and
masses are similar to those of
giant molecular associations (GMAs) found in the nearby 
spiral galaxies M~51, M~100 (Rand 1993a, 1993b, 1995), and NGC~4414 
(\cite{sak96}). However, the clumps in NGC~7479 are kinematically very 
different from those observed in the other galaxies (\S~4.1.2).

\begin{figure*}
\psfig{figure=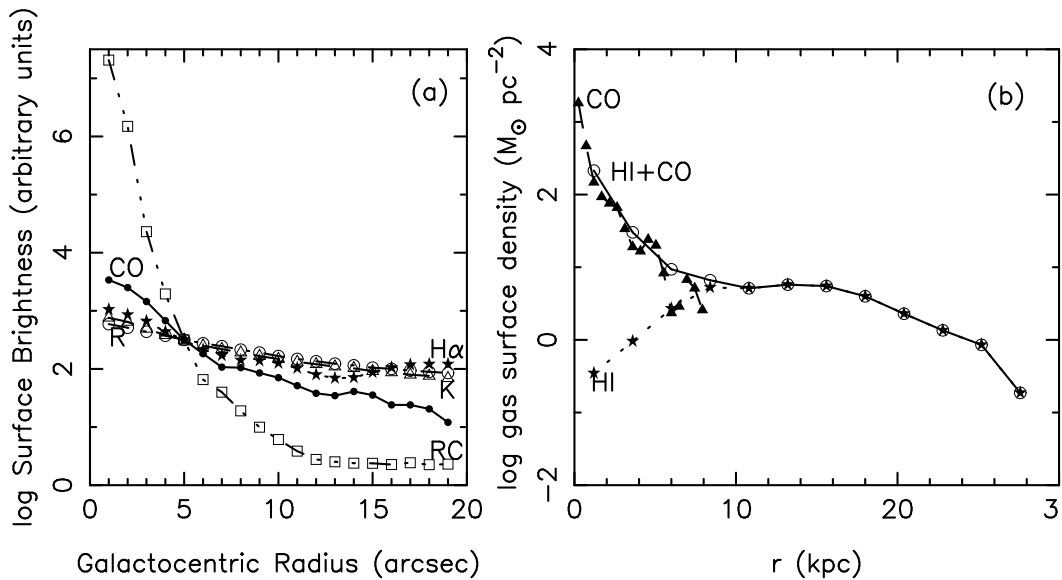,height=3.0in}
\caption{({\it a}). The azimuthally averaged radial 
profiles of the CO
emission, H$\alpha$ emission, $R$-band and $K$-band light, and $\lambda$ =
21~cm radio continuum (RC) in the central area of NGC~7479. The data have 
been smoothed to the same spatial resolution 
of 2\farcs 5, except the radio continuum (4\arcsec). For the CO, the vertical 
axis gives the total molecular mass surface density in
$M_{\sun}$~pc$^{-2}$,
calculated using the standard CO-H$_{2}$ conversion factor. The other
profiles, which have arbitrary units, have been
normalized to agree at 5.0\arcsec. The H$\alpha$ data are from Young \&
Devereux (1991), the $R$-band data were kindly provided by P. Martin and
D. Friedli, taken at the Steward Observatory 2.3~m telescope on Kitt Peak.
The $K$-band data are from Quillen et al. (1995). The 21~cm radio
continuum is from Laine \& Gottesman (1998). ({\it b}) 
Azimuthally
averaged radial profiles of the CO (2\farcs 5~spatial resolution) 
and \ion{H}{1} (15\arcsec~spatial resolution) emission in NGC~7479,
together with the total (molecular + neutral hydrogen) gas profile.
The vertical scale gives the mass surface density in M$_{\sun}$ pc$^{-2}$.
The molecular profile is shown with filled triangles, connected with dashes, 
the \ion{H}{1} profile with stars, connected with dots, and the
total gas (molecular + neutral hydrogen) profile with open circles, connected 
with a solid line. \label{fig5}}
\end{figure*}

The azimuthally averaged radial profile
of the molecular gas is compared to the radial profiles of $R$- and $K$-band 
light, H$\alpha$, \ion{H}{1}, and $\lambda$ = 21~cm radio continuum emission 
in Figure~\ref{fig5}.
The optical and near-infrared images  were first
smoothed to the spatial resolution of our CO data (2\farcs 5). However, the 
radio continuum and the \ion{H}{1} data (from \cite{lai98a}) have resolutions 
of 4\arcsec~and 15\arcsec, respectively.

Within the central 7\arcsec~the CO emission intensity rises by more than
an order of magnitude, as in NGC~3504
(Kenney, Carlstrom, \& Young 1993). Unlike
NGC~3504, the $R$- and $K$-band profiles are shallower than the CO profile
in this area. This may reflect the
lack of an intense starburst in the nucleus of NGC~7479, its 
relatively
small bulge component or large extinction. The CO to H$\alpha$ intensity 
ratio is approximately constant
in the central 2\arcsec, where the nuclear gas disk lies, as in
NGC~3504 (\cite{ken93}). Outside this inner region NGC~3504 and 
NGC~7479 show opposite
behaviours: in NGC~3504 the H$\alpha$ gradient becomes steeper than
the CO intensity profile, whereas in NGC~7479 the opposite is true.
This could mean that the star formation efficiency increases outwards
from the center in NGC~7479. Unfortunately, there are no suitable radio
continuum
measurements which would allow us to assess the effects of extinction.
In other galaxies comparisons between the 6~cm thermal emission and
H$\alpha$ have shown that extinction can change the ``picture'' (e.g.
M 100; Garc\'\i a-Burillo et al. 1998).

The radio continuum
emission is very strongly peaked in the nucleus. These data are
shown at 4\arcsec~resolution, so the gradient at a higher resolution
is likely to be even larger. If the radio continuum is regarded as a
star formation tracer, the star formation efficiency decreases with
radius as in NGC~3504. However, the nucleus of
NGC~7479 is slightly active (LINER), and the radio continuum 
may also have a contribution from the nucleus, not related to star formation.
   
Figure~\ref{fig5}{\it b} shows the striking difference in the distribution of 
\ion{H}{1}
and CO emissions. It should be noted that our CO observations did not
extend very far out into the spiral arms of NGC~7479. The atomic gas 
appears to have been converted into the
molecular form at the higher pressures of the central potential well,
aided by the self-gravity of the gas.

\subsection{Gas motions}
 
The line of sight velocity field (Fig.~\ref{fig3}) is complex and
difficult to interpret. The contours in the nuclear few arcsec region
are aligned perpendicular to the PA of the outer disk 
(22\arcdeg; \cite{lai98a}), consistent with solid body
rotation. Beyond 3\arcsec~radius in the south, the line of sight velocity
contours are aligned roughly along the bar. This morphology of the 
velocity field is consistent with a large radial velocity component (along
the bar; e.g., \cite{lin96}). The northern
part of the bar does not have a sufficiently extended emission region in which
we could estimate the character of the gas motion.

The central 4\arcsec~region of the velocity field where the circularly 
rotating gas and the gas flowing along the bar meet has a 
Z-shape. To better understand the gas
motions in the bar we have made position--velocity plots at several
different PAs and spatial locations.

\begin{figure*}
\psfig{figure=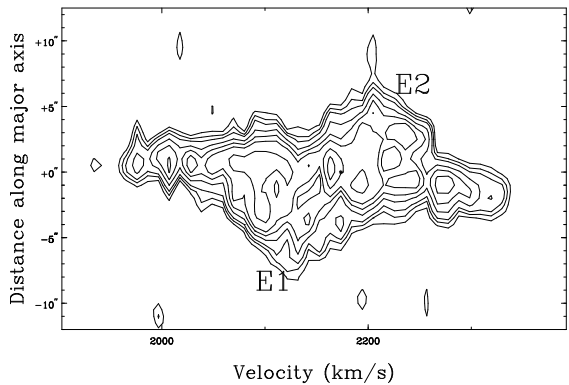,height=2.0in}
\caption{The major axis CO position--velocity plot. The contours 
are at 2, 3, 4, 6, 8, 10, 14, 18, 22, 24 times the rms noise in the
channel maps (see Table~\ref{tbl-2}). The extensions of the emission
towards the south and north have been marked with E1 and E2, respectively.
A possible origin of these features is discussed in the text.\label{fig6}}
\end{figure*}

\begin{figure*}
\psfig{figure=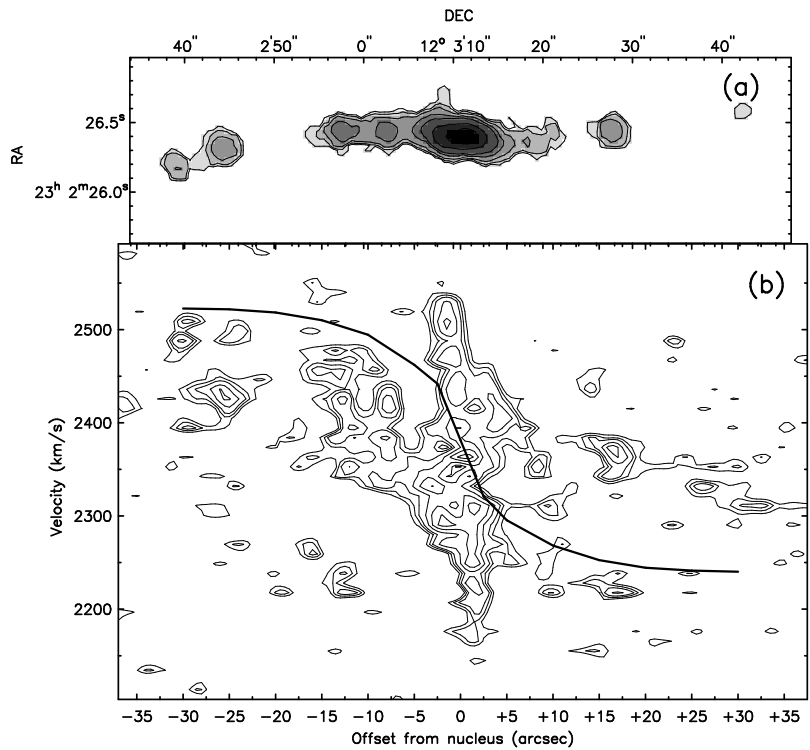,height=4.0in}
\caption{({\it a}). The zeroth moment plot 
(Fig.~\ref{fig2}). ({\it b}). The position--velocity plot along 
the bar at PA $177\arcdeg$ drawn at the same scale in x-axis as ({\it a})
so that the two can be compared easily. The rotation curve derived from the
gravitational potential obtained with the help of a $K$-band image
(Quillen et al. 1995) has been drawn with a solid line in the position--velocity 
plot (after correcting it for the PA and the inclination of the plot)
to show the departures from pure circular motion. \label{fig7}}
\end{figure*}

Position--velocity plots
close to the major axis (at PA 25\arcdeg) and along the bar are presented 
in Figures~\ref{fig6} and \ref{fig7}, respectively. Both plots are centered 
at the nuclear CO emission 
peak (R.A. 23$^{\rm h}$ 02$^{\rm m}$ 26\fs 37, Dec.
12\arcdeg~03\arcmin~11\farcs 1; B1950.0) as
determined from the map of integrated CO emission (Fig.~\ref{fig2}), and 
represent planes (width 0\farcs 5) of the original data cube.  
Outside the 
central 2\arcsec~region, the mean CO velocity within each emission region
along the bar does not deviate 
from the mean velocity (2360~km~s$^{-1}$) of the 
nuclear region by more than 50 km~s$^{-1}$ (Fig.~\ref{fig7}).
Fig.~\ref{fig7} also has a
line which indicates the location where emission would be expected if it 
followed the rotation curve derived from the
gravitational potential, obtained by converting the $K$-band surface
brightness into a mass surface density (\cite{qui95}). Since Quillen et
al. (1995) did not consider the three-dimensional nature of the bulge, and the
{\it M}/{\it L} ratio as determined from a $K$-band image may be variable near the
nucleus, the gravitational potential and the rotation curve in the central 
8\arcsec~are uncertain. 

The remarkable departures of the observed emission from the expected
location traced by the line in Fig.~\ref{fig7} outside the nuclear region
indicate that the gas motion along
the bar is inconsistent with pure circular rotation. Assuming
that the northwestern side of the galaxy is the near side (spiral arms
trailing) and that there are no substantial vertical motions, it is
possible to make
a rough estimate of the radial velocity of the gas along the bar, taking
the tangential speed of the gas in the bar from the bar rotation
rate. There will an additional component of tangential speed, but if the
gas is captured in orbits that are elongated along the bar, a first order
estimate of the tangential velocity is provided by the bar pattern speed.  
Using the bar pattern
speed of 27~km~s$^{-1}$~kpc$^{-1}$ (\cite{lai98b}), the tangential velocity
at the distance of 15\arcsec~(3.2~kpc) is 65~km~s$^{-1}$. The observed
velocity at this point in the clump north of the center is only slightly
higher, 2370~km~s$^{-1}$, than the velocity of
the nucleus. Accounting for the difference between the
directions of the gas bar major axis and the line of nodes (about 31$\arcdeg$)
and the inclination (51$\arcdeg$; \cite{lai98a}), the
observed emission velocities can be reproduced with a radial inflow 
component of 110 km~s$^{-1}$.   

The plot along the CO major axis (Figure~\ref{fig6}{\it a}) also gives a hint of the existence 
of two
separate kinematic components. The first component shows up as a large linear 
velocity gradient
across the central few arc seconds (190~km~s$^{-1}$ arcsec$^{-1}$ or 
1.2~km~s$^{-1}$~pc$^{-1}$) that is consistent with solid body rotation. 
After correcting the observed velocities 
for the inclination, the maximum rotation velocity 
in the nuclear disk is 245 km~s$^{-1}$ at a radius of 1\arcsec. 
The noncircular kinematic component can be seen as extensions toward lower
declinations and lower velocities (marked with E1 in Fig.~\ref{fig6}{\it a}) 
and higher 
declinations and higher velocities (marked with E2 in Fig.~\ref{fig6}{\it a}).

\begin{figure*}
\psfig{figure=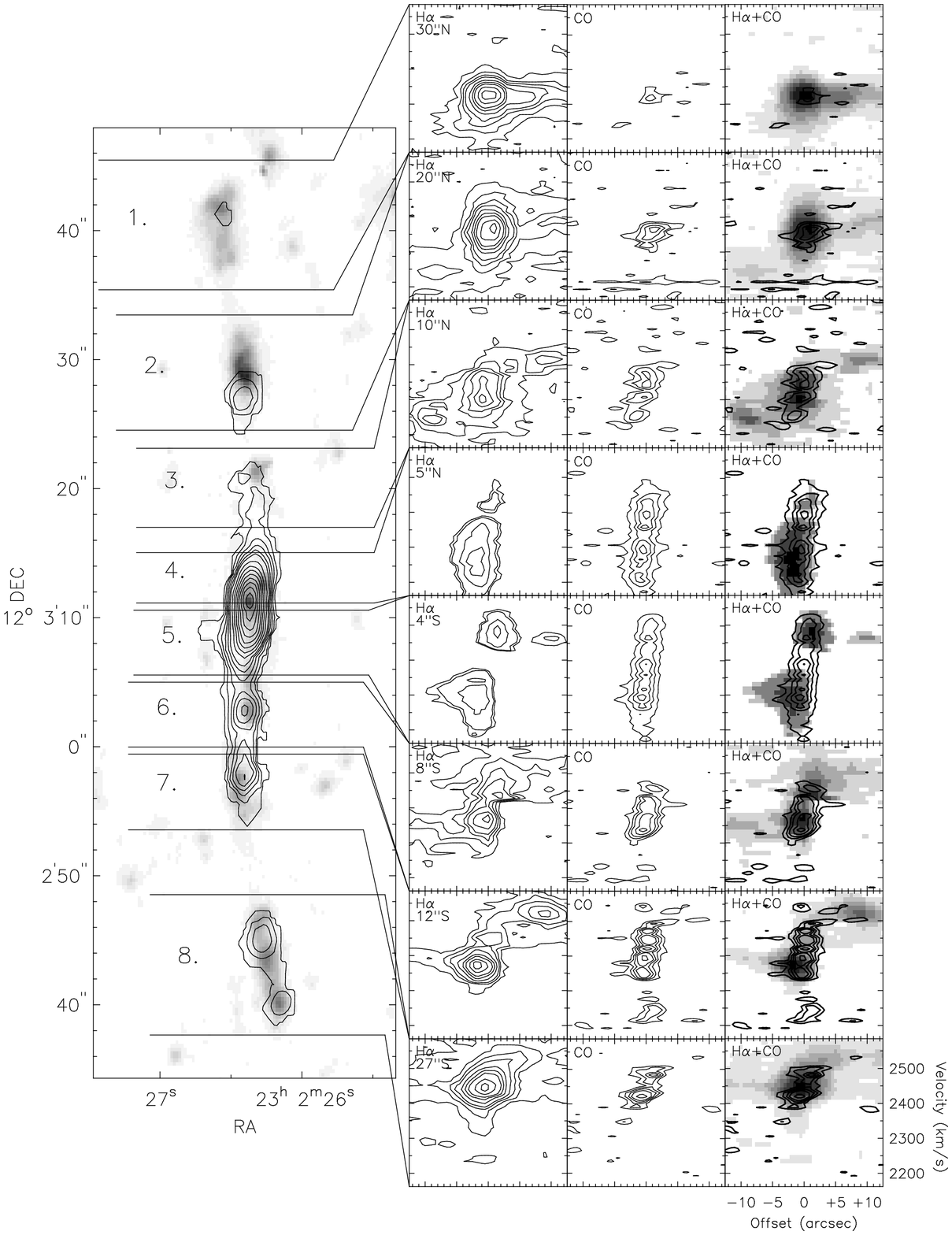,height=6.0in}
\caption{An overlay of the CO emission intensity on
a gray scale image of the H$\alpha$ emission is shown on the left (the 
0\farcs 8~resolution 
H$\alpha$ image was kindly given to us by Dr Johan Knapen). The contours of 
the CO emission are given at 40, 120, 200, 280, 360, 440, 600, 760,
920, 1080, 1240,
1400, 1560, 1720 H$_{2}$ atoms cm$^{-2}$, using the standard conversion
factor between CO intensity and H$_{2}$ surface density. The regions
in which the emissions were integrated for the position-velocity maps
are separated by solid lines, and numbered from 1 to 8. The corresponding
position--velocity maps are shown on the right, the first column giving the
H$\alpha$ maps, the second column the CO maps, and the third column both
overlaid, using gray scales for the H$\alpha$ emission. The velocities are
heliocentric. The contours on the position--velocity maps are fractions of
the peak intensity in a map, chosen to give a clear presentation of the
emission. The strongest CO and H$\alpha$ emission occurs in a narrow
ridge with a large velocity gradient. Weaker H$\alpha$ emission is also
observed both upstream and downstream from this ridge. \label{fig8}}
\end{figure*}

We have also made position--velocity maps perpendicular to the bar (at
PA $90\arcdeg$) at several positions along the bar, using both the CO 
and the H$\alpha$ data
cubes. The resulting maps, together with an image of the integrated 
CO and H$\alpha$
emissions, are shown in Figure~\ref{fig8}. The fluxes between the horizontal boundaries
indicated in the map were summed together to improve the signal-to-noise
ratios within every numbered region. The velocity resolutions
are 10.4~km~s$^{-1}$ (CO) and 12.1~km~s$^{-1}$ (H$\alpha$).


The overlay of the total CO intensity (at 2\farcs 5 spatial resolution) 
contours on an H$\alpha$ image  
(at 0\farcs 8 spatial resolution, kindly given to us by Dr. J. Knapen) 
reveals that the CO and H$\alpha$ emission
peaks occur mostly at the same locations within the available spatial
resolution. However, in regions
3 and 7, the H$\alpha$ emission maxima are at a larger radius and larger 
distance from the bar major axis than the
CO maxima. The position--velocity maps show that in H$\alpha$, two
velocity ``plateaus'' of emission, both upstream and downstream from the
dust lane, are connected by a steep CO (and H$\alpha$) velocity gradient. 
Strong CO emission only exists in the
region of the steep velocity gradient.

The largest velocity gradient in the CO position-velocity maps presented 
in Fig.~\ref{fig8} is about 240~km~s$^{-1}$ arcsec$^{-1}$ in region 7.
Correcting this for the inclination gives a gradient of
1.9~km~s$^{-1}$~pc$^{-1}$. The least
steep gradients further out along the bar have values around 
0.5~km~s$^{-1}$~pc$^{-1}$. 

\section{Discussion}

\subsection{Gas flow and molecular gas in the bars}

\subsubsection{Comparison of observations and models}

Observations of NGC~7479 show dust lanes displaced toward the leading edges 
of the bar, accompanied by CO emission (Fig.~\ref{fig10}) which traces
dense compressed molecular material. The large velocity gradients that we
see in our observations occur along these gas/dust lanes. This is
consistent with the general expectation of shocks in the compressed
interstellar medium. In shocks the component 
of velocity perpendicular to the shock has an abrupt change in magnitude. 
The large velocity gradients in our
observations are also consistent with the models which show that gas 
streaming along the bar has an apogalacticon near the bar dust lanes, where 
the gas flow 
changes from having a radially outward directed velocity component to 
having a radially inward directed component (see, e.g., Fig.~5 in Roberts, 
Huntley,
\& van Albada 1979). The line of sight at which the bar in NGC~7479 is
observed
is intermediate between perpendicular and parallel to the bar, and thus 
favorable for seeing the large gradients in the line of sight velocity 
across the bar dust lanes.

\begin{figure*}
\psfig{figure=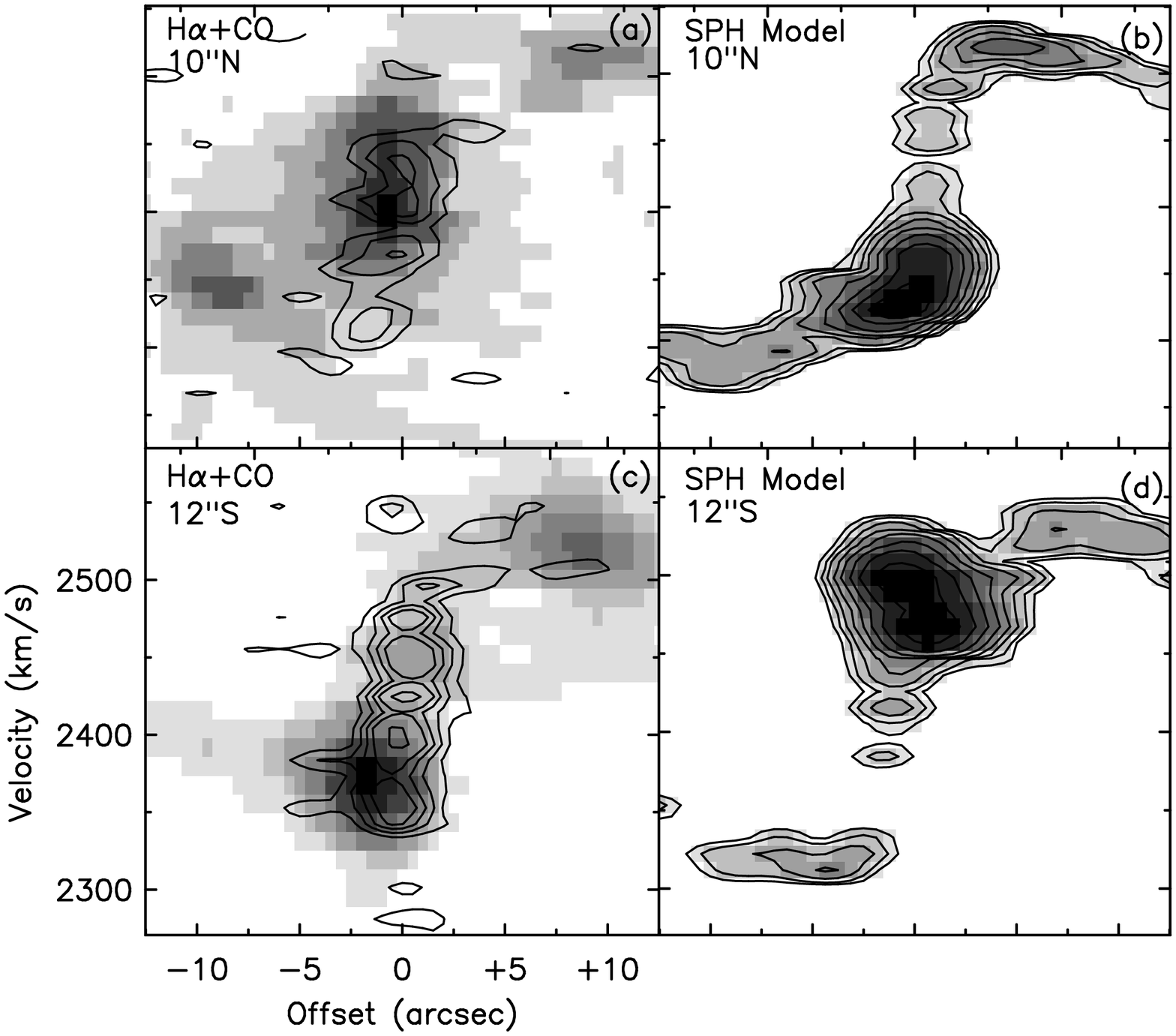,height=6.0in,angle=270}
\caption{({\it a}). Gas 
position--velocity maps of region 3 in Fig.~\ref{fig8}. The H$\alpha$
emission is shown with gray scales and the CO emission with contours.
({\it b}). A position--velocity map of a region corresponding to ({\it a})
in the most successful model in Laine et al. (1998), at {\it
t}~=~9~$\times$~10$^{8}$~yr. ({\it c}). As ({\it a}), but showing region 7
in Fig.~\ref{fig8}. ({\it d}). As ({\it b}), but corresponding to region 7
of Fig.~\ref{fig8}. 
A comparison of observations and models shows that the model has
succeeded in reproducing the observed morphology. Since the shock in the
model takes place along the dust lane, we conclude that the observed
large velocity gradients are consistent with a strong bar shock. \label{fig9}}
\end{figure*}

\begin{figure}
\psfig{figure=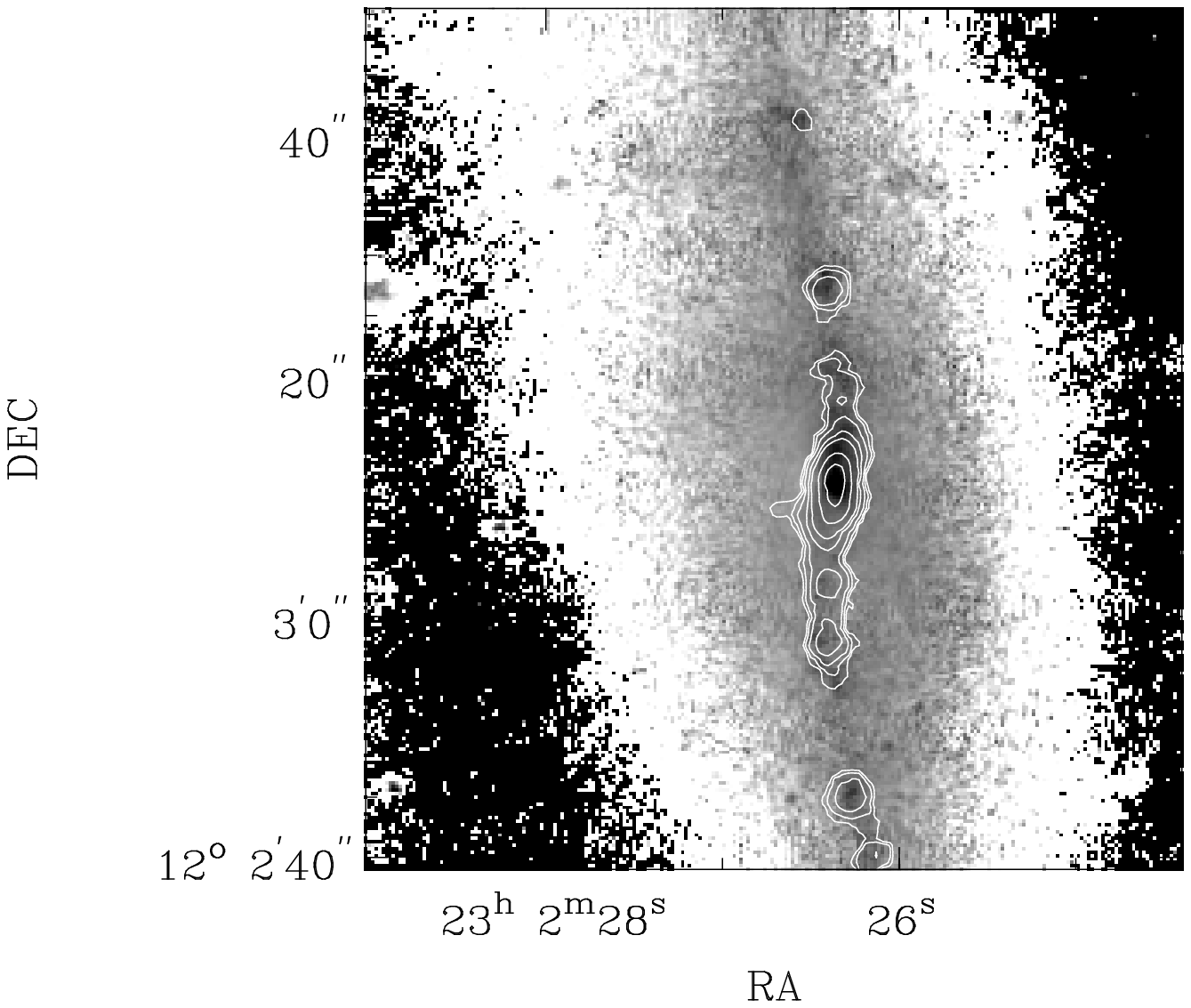,height=2.5in}
\caption{The contours of the naturally weighted CO emission
overlaid on the gray scale image of a $J-K$ color map (Laine 1996), using
data taken at the United Kingdom Infrared Telescope (UKIRT). 
Darker colors are
redder regions in the gray scale map. The contour levels are at
(4, 8, 16, 32, 48, 96, 144) $\times$ 10$^{21}$ H$_{2}$ molecules 
cm$^{-2}$. The excellent spatial correlation shows that strong CO emission
is associated with the high extinction in the bar dust lane. \label{fig10}}
\end{figure}

Figure~\ref{fig9} shows position-velocity maps from observations and
from a Smoothed Particle Hydrodynamics (SPH) bar flow model 
(\cite{lai98b}). The position--velocity slices in both the observations
and the models were taken across the bar at the same
distances from the nucleus, and are thus directly
comparable. The model data were taken from the run with the best fitting 
bar pattern speed (27~km~s$^{-1}$~kpc$^{-1}$).
The two-dimensional simulation frame was projected to the orientation
of NGC~7479 (inclination 51$\arcdeg$, PA 22$\arcdeg$; \cite{lai98a}; 
the angle between the stellar bar and the kinematical major axis in the plane of 
the galaxy 
25$\arcdeg$). The spatial resolution of the model was smoothed by a 
Gaussian to 2\farcs 5 
(400 pc) and the velocity sampled at 10.4 km~s$^{-1}$ bins, comparable to 
the CO (and roughly, the H$\alpha$) observations.

SPH represents total gas, and therefore it traces both the 
compressed regions of dust lanes and dense 
molecular gas as well as the more diffuse ionized gas. In both the models 
and the observations (Figures~\ref{fig7}, \ref{fig8}, \ref{fig9},
\ref{fig10}), the observed velocity changes by 
about 150--200 km s$^{-1}$ across the highest gas 
concentrations (which in optical and near-infrared observations are seen as a 
dust lane; Fig.~\ref{fig10}), but the H$\alpha$ emission ``plateaus'' upstream and 
downstream from the dust lane have relatively small velocity gradients.  
Our kinematical results support the expectation that the bar dust
lane traces shocks in the gas flow. 

The models also show the
velocity ``plateaus'' of gas upstream and downstream from the location of
the large velocity gradient. The
downstream ``plateaus'' suggest that some fraction of the gas has moved
out of the shock region. However, the mean velocity of the shock regions
is consistent with radial inflow motion as mentioned in \S~3.2. Therefore, 
apart from gas that appears to have moved downstream, away from the shock
front, our observations are consistent with the picture given by Regan, Vogel, 
\& Teuben (1997, RVT; see also \cite{reg96t}) who claim that
all gas that encounters the dust lane subsequently flows down the dust 
lane into the nuclear region.

Our observations and the comparison to the SPH model have shown
that the response of gas (including molecular) to the forces along 
the bar in NGC~7479 is more consistent with the behavior of dissipative 
diffuse gas
under hydrodynamic forces (pressure, viscosity) than the response of giant 
dense molecular 
gas clouds which might be expected to react like a ballistic particle. Both
the grid-based ideal gas hydrodynamic codes and SPH codes have  
modeled gas flows, including the compression regions or shocks,
satisfactorily in most cases. ``Sticky
particle'' (SP) codes smear the shocks over relatively large spatial 
scales, depending on the particle size, and may have difficulties modelling the
100 pc scale, steep velocity gradients in the bar dust lane of NGC~7479. It 
is unclear at present  
which code is the most realistic in modeling the behavior of the
interstellar medium (ISM). The equation of state of the ISM is unknown and
the multiphase numerical modeling of the gas has not been done yet. 

\subsubsection{Properties of molecular gas in the bars} 

The star formation and gas flow properties along bars would be easier to
understand if the physical state of the CO emitting regions was known. 
Specifically, we would like to know if the CO emitting regions 
are gravitationally bound.
RVT claim that the excellent agreement between their observations of
NGC~1530 and the hydrodynamic models based on diffuse ideal gas suggests
that the gas they traced with H$\alpha$ kinematics is diffuse gas and
not concentrated in gravitationally bound giant molecular clouds (GMCs;
masses up to a few times 10$^{6}$~$M_{\sun}$) or 
even larger and more massive GMAs, which
might behave more like ballistic particles than diffuse gas. Our
observations and the comparison to models also suggests that the molecular
gas in the bar of NGC~7479 does
not behave like ballistic particles. In the disk of spiral galaxies the 
situation appears different, as complexes of GMAs have been found in
at least the nearby galaxies M~51 (Rand 1993a, 1993b; Vogel, Kulkarni, \& 
Scoville 1988), M~100 (\cite{ran95}), and NGC~4414 (\cite{sak96}). 

The CO luminosities (masses) and diameters of the CO emission regions
along the bar in NGC~7479 are roughly similar to those seen in the {\it
disks} of 
other spirals such as M~51, but the kinematics along the bar of NGC~7479 are 
very different. NGC~7479 has very large velocity gradients of 
1.9~km~s$^{-1}$~pc$^{-1}$
across the emitting regions. Therefore, the CO emission regions along the bar 
are unlikely to be self-gravitating.

The virial masses of the CO emission regions in $M_{\sun}$ can be roughly 
estimated from 

\begin{equation}
M_{{\rm vir}} = 550D(\sigma_{1{\rm d}})^{2}
\end{equation}
(\cite{ran93a}; \cite{sco87a}). Here $D$ is the observed FWHM diameter in
parsecs, and $\sigma$$_{1d}$ is the one-dimensional velocity dispersion in
km~s$^{-1}$.
This formula assumes a $1/r$ density profile for the gas within a GMA.
The mass required for the GMA to be self-gravitating is half of this
mass. Using a velocity dispersion of 150 km~s$^{-1}$ and a FWHM diameter
of 150 pc (about 1\arcsec), the resulting virial mass is about 
2~$\times$~10$^{9}$~$M$$_{\sun}$. For comparison, the observed masses of
the clumps along the bar (assuming the standard conversion factor) range
from $9 \times 10^{6}$~$M_{\sun}$ to $2 \times$ 10$^{8}$~$M_{\sun}$.
Thus it is unlikely that the CO gas
concentrations that lie along the bar are virialized or even
gravitationally bound. 

\subsection{Star formation in the bar of NGC~7479}

Star formation rates (SFRs) from the H$\alpha$ fluxes in the offset gas 
and dust lane along the bar have been calculated from
\begin{equation}
{\rm SFR_{H\alpha}} = 7.07 \times 10^{-42}L({\rm H\alpha})~M_{\sun}~{\rm
yr^{-1}},
\end{equation}
(\cite{bus87}; see also \cite{kcut}). Equation (3) assumes
a Salpeter initial mass function (IMF) down to stars
of mass 0.1~$M_{\sun}$ and up to stars with masses of 100~$M_{\sun}$. The 
H$\alpha$
fluxes have not been corrected for extinction, so the SFRs from equation
(3) are lower limits. The
integrated SFR in the bar region is around 0.2 $M_{\sun}$ yr$^{-1}$. 
Uncertainties due to extinction and the choice of the IMF with its cutoffs
are discussed in Bushouse (1987). Specifically, Bushouse compared the global 
star formation
rates as calculated from the H$\alpha$ fluxes (uncorrected for extinction
effects) to global far-infrared fluxes. The latter were
found to be three times larger on average, and therefore an average extinction
correction factor of three applies to isolated galaxies. For comparison,
Martin~\& Friedli (1997) measured the ratio of the H$\alpha$ and H$\beta$
lines in a few selected \ion{H}{2} regions along the bar, and obtained
a bar extinction value of 0.8 mag in H$\alpha$, which corresponds to a
factor of 2.
Multiplying the measured H$\alpha$ flux by 2.5 results in
a star formation rate [SFR(H$\alpha$)] of  
0.5~$M_{\sun}$~yr$^{-1}$ along the bar of NGC~7479, comparable
to the value (0.4 $M_{\sun}$ yr$^{-1}$) obtained by Martin~\& Friedli
(1997). The SFR for the whole galaxy, calculated from the
infrared luminosity [$L$(IR); 8.9$\times$~10$^{10}$ L$_{\sun}$;
\cite{you89}], is about 45 $M_{\sun}$ yr$^{-1}$ 
[SFR(IR)~=~1.34x10$^{-43}$ $L$(IR); \cite{hunt}; \cite{bus87}],
90 times larger than the SFR along the bar and thus largely not seen in 
optical tracers, although the H$\alpha$ luminosity of the bar region is 
only about 10$\%$ of the total H$\alpha$ luminosity of NGC~7479.

Bushouse (1987) finds that SFR(IR) is 3--6 times larger than SFR(H$\alpha$)
among normal galaxies, (3 times in isolated, 6 times in interacting
galaxies), most likely because of extinction in the optical.
Yun~\& Hibbard (1998) confirmed the significantly larger SFR(IR) values 
and also found that the H$\alpha$ component seen
in the nuclear starburst systems is dominated by the starburst superwind
and scattered emission.
It is possible that the nucleus of NGC~7479 hosts 
a starburst behind dozens of magnitudes of optical extinction.
About 15$\%$ of the 21 cm continuum emission is concentrated in the
central beam (4\farcs 3 $\times$ 3\farcs 85), giving a rough size scale
for the starburst region (\cite{lai98a}).

The large velocity gradient of 0.5--1.9~km~s$^{-1}$~pc$^{-1}$
across the bar dust lane of NGC~7479 
is likely to be dynamically important and affect star formation.
The large velocity gradient can prevent large gas concentrations
within the bar dust lane from being self-gravitating, as indicated by
our simple calculation in \S~4.1.2.

The kinematical conditions in the bar dust lane are quite different from 
those in the
spiral arms of M~51, which has unusually strong spiral density waves
with small pitch angles.
Although there is a density wave induced velocity gradient 
of 0.05--0.1 km~s$^{-1}$~pc$^{-1}$ across the front half of the spiral
arms of M 51 (\cite{vog88}; \cite{ran93a}),
this gradient is an order of magnitude smaller than that in the
bar of NGC~7479, and is furthermore in a direction which reduces the 
amount of shear which otherwise occurs in a disk with a flat
rotation curve.

Thus, in spiral arms the largest concentration of dense gas  
and the regions of reduced shear occur in the same places, and both effects 
increase the susceptibility of gas to gravitational collapse.
In bars the largest concentration of dense gas occurs 
in a region of a large velocity gradient, and this opposes the
susceptibility of gas to gravitational collapse.

This could be the reason why star formation is observed upstream and
downstream from the dust lane as well as within it.
Although the regions upstream and downstream from the
shockfront have lower gas density, their smaller velocity gradients may 
enable star formation. As discussed in \S~3, the regions upstream and
downstream may have molecular gas that has a column density which is too
low to be detected when averaged over the synthesized beam, but the
density may be high enough to enable star formation, making it
easier to understand the wide distribution of H$\alpha$ emission across
the bar.

The regions with the highest gas density are regions with the
largest velocity gradients. 
Although stars are certainly forming in this dense dust lane gas,
the star formation rate may be limited by the large velocity gradients:
the long timescale for star formation along the bar
(2~$\times$~10$^{9}$~$M_{\sun}$/0.5~$M_{\sun}$~yr$^{-1}$~=
4.0~$\times$~10$^{9}$ yr) 
implies that the bar gas is not experiencing a starburst. Again, this
conclusion is subject to the effects of extinction, as discussed before,
although these effects are unlikely to change the basic conclusion.
Elmegreen (1979) and Kenney \& Lord (1991) have also suggested
that the tidal forces on the gas clouds along the bar dust lanes
may inhibit star formation.

\subsection{Nuclear gas mass fraction}

The molecular gas mass in the centers of some spiral galaxies is a large 
fraction (up to at least 50$\%$) of the
total dynamical mass (\cite{tur94}; \cite{ken97a}). Under these
conditions, the self--gravity of the molecular gas becomes important and
may drive further inflow towards the center (\cite{shlo89}; 
\cite{shlo90}; \cite{hel94}; \cite{kna95}; \cite{wad92}). The gas mass 
fraction in the central disk of NGC~7479 can be
calculated in the region that exhibits circular rotation,
using the standard CO luminosity to H$_{2}$ mass
conversion factor.
The rotation velocity in the center is about 245 km~s$^{-1}$.
We use a radius of 1\arcsec~for the nuclear gas disk.
Using a distance of 32 Mpc, the total dynamical central mass becomes
\begin{equation}
M_{{\rm dyn}}~=\frac{rv^{2}}{G}~=2.2~\times 10^{9} M_{\sun}.
\end{equation}
 
The molecular mass in this area is $3.3~\times$ 10$^{8}$ $M_{\sun}$ or
$15\%$ of the dynamical mass. 
The molecular gas mass is a considerable part of the total
mass in the center, although it does not dominate the dynamics. If the rest of 
the mass is
stellar, the average mass-to-luminosity ratio in the $K$-band in this central
area is 0.7 in solar units. This would imply a relatively young
(a few Gyrs) stellar population or alternatively a metal rich old
population (\cite{wor94}).

Since the star formation rate along the bar is 0.5~$M_{\sun}$ yr$^{-1}$
and the gas inflow rate is at least 
4~$M_{\sun}$~yr$^{-1}$ (\cite{qui95}; \cite{lai96}), most of the  
gas presently in the bar region should reach the 
circumnuclear region before it turns into stars. The uncertainties are caused 
by the unknown amount of extinction, which
may severely suppress the observed H$\alpha$ fluxes along the bar dust
lane and cause an underestimate of the bar SFR. The net inflow of
2~$\times~10^{9}$~$M_{\sun}$ of gas would double the nuclear mass in less
than a Gyr, meaning that NGC~7479 is in a short-lived evolutionary state.
The resulting increase in the nuclear mass would likely drive the evolution
of the galaxy towards an earlier Hubble type (\cite{kor93}; \cite{fri93}; 
\cite{ken97}).

\section{Conclusions}

The high spatial resolution CO observations of NGC~7479 together with 
H$\alpha$ data have shown the 
gas distribution and motions in
a dynamically young barred spiral galaxy (cf. \cite{ken97}; 
\cite{lai98c}). Our 
main conclusions are summarized in the following:

1. The CO emission has a strong central peak, but about $50\%$ of
the emission comes from regions outside the nuclear zone, mostly from 
a clumpy distribution along the bar dust lane.

2. The bar dust lane and the CO gas emission occur at the same locations 
within the resolution of the observations.

3. The CO velocity field is consistent with circular rotation only in the
central 2\arcsec~(300~pc) diameter region. Outside this region, the velocity 
field deviates strongly from pure rotation
and the velocities must have a substantial (at least 100 km~s$^{-1}$) radial 
component along the bar.

4. The CO emission across the bar dust lane has a large velocity gradient 
(up to 1.9~km~s$^{-1}$~pc$^{-1}$) with a projected velocity change up to
200 km~s$^{-1}$. The H$\alpha$ emission shows a similar 
large gradient, but it also has ``plateaus'' of constant velocity emission
both upstream and downstream from the large velocity gradient. The 
large velocity gradient is consistent with combined effects of a shock and 
a change in the radial direction of gas streamlines in the bar dust lane.

5. The gas clumps along the bar have masses from 
9$\times$10$^{6}$~$M_{\sun}$ to 2$\times$10$^{8}$~$M_{\sun}$ if the
standard CO--to--H$_{2}$ conversion factor is assumed. However, the large 
velocity gradient across these complexes in the bar dust lane makes this 
assumption suspect.

6. No strong CO emission exists in the bar region outside the dust lane, 
and the gas complexes in the dust lane are unlikely to be gravitationally 
bound because of the large velocity gradients across them.  

7. Based on the relative strengths of the CO and H$\alpha$ emissions in the
nucleus and along the bar, we conclude that either the optical emission from 
the nucleus is strongly reduced
by extinction or the star formation efficiency in the nucleus is
lower than in regions along the bar. 

8. The star formation timescale along the bar is long, a few Gyr.
The large velocity gradients in the bar dust lane may prevent the
bar gas from forming stars more rapidly.

9. Along the bar, the gas inflow rate of 4~$M_{\sun}$~yr$^{-1}$ is much 
greater than the star formation rate of 0.5~$M_{\sun}$ yr$^{-1}$, implying 
that most of the gas presently in the bar will reach the circumnuclear region.
If that happens, within the next $\sim$~1 Gyr the central mass of NGC~7479
should roughly double. However, extinction in the bar dust lanes and in
the nuclear region causes the cited SF rate to be a lower limit, and the
actual amount of gas reaching the circumnuclear region is likely to lower.

10. The nuclear gas mass fraction is likely to be at least $15\%$ of 
the total
dynamical mass within the central gas disk, and therefore, the molecular gas
affects the dynamics in the nuclear disk.

\acknowledgments

We are very grateful to Dr Nicholas Z. Scoville for financial support for the
publication of our results and the
observing run. We are grateful to Drs Michael Regan and Stuart Vogel for 
providing us
with the H$\alpha$ data for comparison with our CO data.  We thank Drs P. 
Martin, D. Friedli, A. Quillen, and J. Knapen for providing most of the 
optical and near-infrared wide-band and H$\alpha$ images used in this work. 
The United Kingdom Infrared Telescope is operated by the Joint 
Astronomy Centre on behalf of the U.K. Particle Physics and Astronomy 
Research Council. Some of the data reported here were obtained
as part of the UKIRT Service Programme. S. L. acknowledges DSR at the
University of Florida for financial support and Dr I. Shlosman for
helpful discussions on bar dynamics. We thank the anonymous referee for
comments that improved the paper. This work was supported in part by
NSF Grant AST 93-14079.

\end{document}